\documentstyle[prd,aps,floats]{revtex}

\begin{document}
\preprint{SUSX-TH-98-004, SUSSEX-AST 98/3-2, PU-RCG/98-3, gr-qc/9803070}
\draft

%
%
\input epsf
\renewcommand{\topfraction}{0.8}
\renewcommand{\bottomfraction}{0.8}
\twocolumn[\hsize\textwidth\columnwidth\hsize\csname 
@twocolumnfalse\endcsname

\title{Black holes and gravitational waves in string cosmology} 
\author{Edmund J.~Copeland}
\address{Centre for Theoretical Physics, University of Sussex, Falmer, 
Brighton BN1 9QJ,~~~U.~K.}  
\author{Andrew R.~Liddle}
\address{Astronomy Centre, University of Sussex, Falmer, Brighton BN1
9QJ,~~~U.~K.}  
\author{James E.~Lidsey}
\address{Astronomy Centre and Centre for Theoretical Physics, University of 
Sussex, Falmer, Brighton BN1
9QJ,~~~U.~K.}  
\author{David Wands}
\address{School of Computer Science and Mathematics, University of 
Portsmouth, Portsmouth PO1 2EG,~~~U.~K.}  
\date{\today} 
\maketitle
\begin{abstract}
Pre--big bang models of inflation based on string cosmology
produce a stochastic gravitational wave background whose spectrum 
grows with decreasing wavelength, and which may be detectable using 
interferometers such as LIGO. We point out that the gravitational wave 
spectrum is closely tied to the density perturbation spectrum, and that the
condition for producing observable gravitational waves is very similar to 
that for producing an observable density of primordial black holes. Detection 
of both would provide strong support to the string cosmology scenario.
\end{abstract}

\pacs{PACS numbers: 98.80.Cq \\Preprint SUSX-TH-98-004, SUSSEX-AST 
98/3-2, PU-RCG/98-3, gr-qc/9803070}

\vskip2pc]


\section{Introduction}

The pre--big bang 
string cosmology scenario is a novel way of producing inflation which
capitalizes on the kinetic energy of a scalar field, the dilaton, rather than
the potential energy as in conventional models \cite{pbb}.  It possesses two
phases, the first known as the dilaton phase and the second the
string phase.  (For a recent review, see e.g.~Ref.~\cite{pbbreview}).
During the dilaton phase, the space--time curvature and gravitational
coupling both grow with time until the former reaches the string scale,
although the latter may still be small.  At this stage non--perturbative
effects become important and the universe enters the string phase, where the
dynamics are much less certain.  Eventually the string phase gives way to the 
standard hot big bang picture.

Scalar (density) and tensor (gravitational wave) perturbations are generated
in the universe during the dilaton phase, and can be calculated using
standard techniques \cite{phen,rise}.  A much-advertized prediction of the
dilaton phase is that the spectrum of gravitational waves is steeply rising
to short scales, in contrast to potential-driven inflation models where it
must decrease as one looks to shorter scales.  In the latter case, the Cosmic
Background Explorer (COBE) observations already guarantee that the stochastic
gravitational wave background lies orders of magnitude below the sensitivity
of even advanced versions of the Laser Interferometric Gravitational Wave
Observatory (LIGO) currently under construction \cite{nosee}.  In the dilaton
phase, however, the spectrum rises as $k^3$ ($k$ being the comoving
wavenumber), which has led several authors to suggest that LIGO may be able
to detect these perturbations \cite{rise}.

However, the adiabatic density perturbations that are also produced
during the dilaton phase have an extremely similar amplitude on a
given scale to that of the gravitational waves \cite{bggmv,bh}.  We
define the scalar and tensor amplitudes $A_{{\rm S}}^2$ and $A_{{\rm
T}}^2$ as in Ref.~\cite{LLKCBA} (note that $A_{{\rm S}}$ is the same
as $\delta_{{\rm H}}$ of Ref.~\cite{LL93} and represents the density
contrast at Hubble--radius--crossing during a matter--dominated era).
The present energy density of gravitational waves is given from the
initial amplitude by \cite{gravom} 
\begin{equation}
\label{omegagw}
\Omega_{{\rm gw}}(k) = \frac{25}{6} \frac{A_{{\rm T}}^2}{z_{{\rm eq}}} \,,
\end{equation}
where $z_{{\rm eq}} = 24\,000 \, \Omega_0 h^2$ is the redshift of matter 
radiation equality. Here $\Omega_0$ and $h$ are the present density parameter 
and Hubble parameter, in the usual units.

We write the tensor to scalar ratio as
\begin{equation}
\label{consistency}
\frac{A_{{\rm T}}^2}{A_{{\rm S}}^2} = \epsilon \,.
\end{equation}
In a conventional inflation model $\epsilon$ is the usual slow-roll
parameter \cite{LL92}, and in the slow--roll
approximation it is bounded, $0 < \epsilon < 1$, and normally much less than 
one.

In the dilaton phase of string cosmology, however, $\epsilon$ equals~$3$.
Although the dilaton phase is far from the usual slow-roll limit, the general
relativistic result that $\epsilon=1/p$ for power-law inflation/deflation,
where the scale factor $a\propto t^p$, still holds.  This result is exact
because the scalar and tensor perturbations obey the same evolution equation
and the ratio is then fixed by their normalization as adiabatic vacuum
fluctuations on small scales.  In conventional power-law inflation $p>1$ and
thus $\epsilon<1$, but in the Einstein frame of low-energy string theory the
dilaton phase corresponds to a collapsing universe with
$p=1/3$~\cite{deflation}.  This is a generic prediction for adiabatic density
perturbations in any model which is conformally equivalent to a collapsing
universe in Einstein gravity, as this represents massless fields with a
maximally stiff equation of state dominating the energy density as the scale
factor $a\to0$.  We will discuss the possible effect of non-adiabatic
perturbations later.

Combining Eqs.~(\ref{omegagw}) and (\ref{consistency}), 
we find that on {\em any} scale~$k$
\begin{equation}
\label{ratio}
A_{{\rm S}}^2 = \frac{1}{3} \, A_{{\rm T}}^2 = 2 \times 10^{-3} 
	\frac{\Omega_{{\rm gw}}}{10^{-6}}  \, \Omega_0 h^2 \,.
\end{equation}
Thus, both $A_{{\rm S}}^2$ and $A_{{\rm T}}^2$ exhibit an increase as
$k^3$ with wave\-number in the pre-big bang scenario \cite{bggmv}.

\section{Gravitational waves}

A very detailed analysis of the detectability of the gravitational waves by
LIGO has been made by Allen and Brustein \cite{ab}.  
LIGO is sensitive to frequencies around $f \approx 100{\rm Hz}$. 
During the dilaton phase
$\Omega_{{\rm gw}}$ grows as $k^3$, and this portion of the spectrum is
characterized by the frequency, $f_{\rm s}$, and fractional energy density,
$\Omega^{\rm s}_{\rm gw}$, at the point where the dilaton phase ends. Note 
that the frequency $f$ and wavenumber $k$ are interchangeable, since we set 
$c=1$. If the string phase is inflationary, then higher frequency 
gravitational waves will be produced, but our
understanding of the generation of perturbations is much less certain.  Allen
and Brustein take the spectrum to have an arbitrary slope $\beta$ in this
region \cite{ab}.  During an inflationary string phase, all scalar and tensor 
perturbations that exited during the dilaton phase will remain beyond the 
Hubble radius.  Thus, in what follows, we assume that Eq.~(\ref{ratio}) 
remains
valid over those scales where $f < f_{\rm s}$ and, furthermore, 
that the frequencies accessible to LIGO lie in this regime, i.e., 
that these modes exited the Hubble radius during the dilaton  phase. 

Allen and Brustein demonstrate that much of the parameter space where
a stochastic gravitational wave background could be detectable by the
initial LIGO configuration is already excluded by primordial
nucleosynthesis bounds \cite{ab,nuc}. {}From here on, therefore, we
focus on the advanced LIGO configuration. For a frequency at the end
of the dilaton phase around $100$ Hz, advanced LIGO can probe to
$\Omega_{{\rm gw}}^{{\rm s}} \sim 10^{-9}$. For comparison, the 
nucleosynthesis bound is $\Omega_{{\rm gw}} \lesssim 5 \times 10^{-5}$.

\section{Density perturbations and black holes}

Density perturbations whose amplitude is of order unity when they
re-enter the Hubble radius can immediately collapse to form primordial
black holes. Because black holes redshift more slowly than the
radiation, which is assumed dominant, even a very modest initial
fraction (perhaps $10^{-20}$ by mass) can be observationally
constrained. Hence, any black holes which form correspond to
high-sigma fluctuations in the density field, whose mean square perturbation
must therefore be well below unity.

Assuming the standard cosmology (we shall examine alternatives later), the 
epoch during the radiation--dominated era when a comoving scale 
$f_*$ equals the Hubble scale is determined by
\begin{equation}
\label{fs}
\frac{f_*}{f_0} = \frac{H_* a_*}{H_0 a_0} 
	\approx \frac{T_*}{T_{{\rm eq}}} \, z_{{\rm eq}}^{1/2}
\end{equation}
where $f_0 = a_0 H_0 = 3h \times 10^{-18}$ Hz is the mode that is just 
re-entering the Hubble radius today. 
Since $T_{\rm eq} = 24\,000\,\Omega_0 h^2 \, T_0 \approx 1{\rm eV}$, 
we have 
\begin{equation}
{f_* \over 100\, {\rm Hz}} \approx {T_* \over 10^9\, {\rm GeV}} \ .
\end{equation} 

The mass of black holes forming from perturbations that collapse 
immediately after re-entry is given by the horizon mass at that time, up to 
a numerical factor of order unity. In a radiation-dominated universe, this 
is given approximately by $M_{\rm hor} \approx 10^{32} (T/{\rm GeV})^{-2} 
{\rm g}$, and the black hole mass for a given mode $f_*$ is therefore 
\begin{equation}
\label{PBHmass}
M  \approx 10^{14} \left( \frac{100 \, {\rm Hz}}{f_*} \right)^2 {\rm g} \,.
\end{equation}

Whether or not black holes of mass $M$ form is governed by the dispersion 
$\sigma$ of the matter distribution smoothed on the length scale $R$ giving 
that horizon mass. The dispersion is defined as (see e.g.~Ref.~\cite{LL93})
\begin{equation}
\sigma^2(R,t) = \left(\frac{10}{9}\right)^2 
	\int_0^{\infty} \left( \frac{k}{aH} \right)^4 
	A_{{\rm S}}^2(k) \, W^2(kR) \, \frac{dk}{k} \,,
\end{equation}
where $A_{{\rm S}}$ is related to $\Omega_{{\rm gw}}$ by Eq.~(\ref{ratio}), 
the time-dependence is carried by the $aH$ factor, and the prefactor appears 
because we are considering radiation domination rather than the usual matter 
domination. We take the smoothing window $W(kR)$ to be a gaussian; for an 
$A_{{\rm S}}^2 \propto k^3$ spectrum the top-hat filtered 
dispersion remains dominated by the shortest scales rather than the smoothing 
scale and so such smoothing is unsuitable.

We first assume that there are no scalar 
perturbations generated during the string 
phase (the `dilaton only' scenario in the language of Allen and Brustein 
\cite{ab}), so the spectrum vanishes for $f > f_{{\rm s}}$. The 
steeply-rising spectrum will guarantee that only modes close 
to $f_{{\rm s}}$ can give significant black hole production.
We measure $R$ in 
units of $k_{{\rm s}}^{-1}$, and consider the dispersion $\sigma_{{\rm hor}}$ 
when that scale $R$ crosses the horizon, so that $aH = 1/R$. Substituting in 
from Eq.~(\ref{ratio}) gives 
\begin{eqnarray}
\label{sigeq}
\sigma_{{\rm hor}}^2(k_{{\rm s}} R) & = & 2 \times 10^{-3} \, \Omega_0 h^2
	\frac{\Omega_{{\rm gw}}^{{\rm s}}}{10^{-6}} \\
 && \quad \times (k_{{\rm s}} R)^4 \, \int_0^1 \tilde{k}^6 W^2(\tilde{k} 
 	k_{{\rm s}} R) \, d\tilde{k} \,.  \nonumber
\end{eqnarray}
For $k_{{\rm s}} R \ll 1$ (small scales) this is small due to the prefactor, 
as the perturbations contributing to $\sigma$ are on longer scales than the 
horizon and have not had 
time to grow to their horizon-crossing value. For $k_{{\rm s}} R \gg 1$ 
(large scales) this is small as the 
dominant short-scale perturbations have been smoothed out. Therefore 
$\sigma_{{\rm hor}}$ peaks for $R \simeq k_{{\rm s}}^{-1}$, and it is on this 
scale that black holes predominantly form.

There are some uncertainties in the exact parameters required for
black hole formation, though these are not particularly important for
our calculations. The usual criterion during radiation domination is
that black holes form in any region with density contrast greater than
a threshold $\delta_{{\rm c}} = 1/3$ when they enter the horizon, and
that the corresponding black hole mass is 0.2 times the horizon mass
(see e.g.~Ref.~\cite{carr}).

Given the dispersion $\sigma$, the fraction of the Universe in regions with 
density contrast exceeding $\delta_{{\rm c}}$ is given by the integral over 
the tail of the 
gaussian, yielding a mass fraction
\begin{equation}
\label{beta}
\beta = {\rm erfc} \left( \frac{\delta_{{\rm c}}}{\sqrt{2} \,
	\sigma_{{\rm hor}}(k_{{\rm s}} R)}  \right) \,,
\end{equation}
where `erfc' is the complementary error function
\begin{equation}
{\rm erfc}(x) \equiv \frac{2}{\sqrt{\pi}} \int_x^{\infty}\exp(-u^2) \, du \,.
\end{equation}
This expression is familiar from Press--Schechter theory in large-scale 
structure studies, and gives the fraction of the total mass 
in black holes with a mass greater than or equal to the 
smoothing mass.\footnote{We have included the factor two multiplier on the 
right-hand side on Eq.~(\ref{beta}) which is added in the large-scale 
structure context to ensure underdense regions contribute to 
gravitationally-bound objects. Inclusion or otherwise of this factor has 
a completely negligible impact on our results.}

Black holes of mass $10^9$g evaporate around nucleosynthesis, while those of 
mass $5 \times 10^{14}$g are evaporating at the present epoch. Black holes 
within this mass range are constrained by a range of different observations, 
summarized in Refs.~\cite{cgl,GL}. Those of mass above $5 \times 10^{14}$g 
have negligible evaporation, and are constrained by their contribution to the 
present total matter density. In all cases, the initial mass fraction of 
black holes must be tiny, since it grows in proportion 
to the scale factor during the long radiation-dominated epoch.

Equation (\ref{PBHmass}) shows that primordial black holes which are 
evaporating
at the present epoch are formed from density perturbations with the same
comoving wavelength as the gravitational waves which LIGO hopes to detect.
As the error function depends so strongly on its argument, we can adopt an
extremely qualitative view of the observations; namely, that for the standard
cosmology the initial mass fraction $\beta$ should be no more than $10^{-20}$
on those scales \cite{cgl,GL}.  As ${\rm erfc}^{-1}(10^{-20}) \simeq 6$ the
observational constraint\footnote{For comparison ${\rm erfc}^{-1}(10^{-10})
\simeq 4.5$, from which we realize that it doesn't really matter what mass
fraction we adopt as the constraint.}  corresponds to $\sigma_{{\rm hor}} <
0.04$.

To convert this into a constraint on $\Omega_{{\rm gw}}^{{\rm s}}$, note that 
for a gaussian smoothing, $W(y) = \exp (-y^2/2)$,
\begin{equation}
\max \left\{ x^4 \int_0^1 \tilde{k}^6 W^2(\tilde{k} x) \, d\tilde{k}
	\right\} \simeq 0.15 \quad {\rm at} \; x \simeq 1.74 \,. 
\end{equation}
Thus, the main black hole formation corresponds to $R = 1.74/k_{{\rm s}}$, 
and 
the fraction of the total mass in black holes above the corresponding 
formation mass (in 
practice dominated by black holes 
close to this mass) is given by 
Eq.~(\ref{beta}). Hence the gravitational wave amplitude leading to a black 
hole density at the current observational limit is, from Eq.~(\ref{sigeq}),
\begin{equation}
\label{Omegagw}
\Omega_{{\rm gw}}^{{\rm s}} = \frac{5 \times 10^{-6}}{\Omega_0 h^2} \,.
\end{equation}
For plausible values of $\Omega_0 h^2$, this is a little below the bound on 
gravitational waves from nucleosynthesis, which in this dilaton-only scenario 
is $\Omega_{{\rm gw}}^{{\rm s}} < 5 \times 10^{-5}$ \cite{ab,nuc}. 

\section{Complications}

\subsection{Non-adiabatic perturbations}

In taking the ratio $\epsilon$ between the tensor and scalar
perturbations to be exactly $3$, we have assumed that the scalar
curvature perturbations are due to purely adiabatic fluctuations. This
is a natural assumption in many conventional models of inflation where
fluctuations in only one scalar field determine the final amplitude of
density perturbations. However, in the low energy effective action
there are many massless fields, with associated spectra of
fluctuations \cite{axions}. This may be very important for calculating
the perturbations on large scales where the fluctuations in the
dilaton, and other fields minimally coupled in the Einstein frame, are
strongly suppressed. Indeed, if the pre--big bang era is to be able to
generate seed perturbations for large-scale structure, then non-adiabatic
perturbations in other scalar fields, such as the axion fields, must
play a significant role \cite{axions}.

A gaussian spectrum of non-adiabatic perturbations, uncorrelated with
the original adiabatic spectrum, can only add to the
overall scalar curvature perturbation power
spectrum~\cite{nonadiabatic,SS}. 
We can represent their effect by introducing an
effective value of $\epsilon_{\rm eff}<3$ during the pre--big bang phase.
The maximum
density of gravitational waves compatible with current limits on the
number density of black holes given in Eq.~(\ref{Omegagw})
then becomes
\begin{equation}
\Omega_{{\rm gw}}^{{\rm s}} = \left(\frac{\epsilon_{\rm eff}}{3}\right)
 \frac{5 \times 10^{-6}}{\Omega_0 h^2} \,.
\end{equation}
That is, non-adiabatic perturbations lower the gravitational wave amplitude 
corresponding to the black hole limits.

\subsection{String phase}

We are assuming that the string phase has just the right properties to place
the end of the dilaton phase into the observable window.  Assuming efficient
reheating ($T_{{\rm reh}} \sim 10^{18}$ GeV), this requires the string phase 
to be inflationary, with an
expansion factor of about $T_{{\rm reh}}/10^{9} \, {\rm GeV} \sim 10^9$, 
since 
during radiation domination $aH \propto T$.  The string phase must be 
inflationary for the gravitational waves generated in the dilaton phase to be
detectable by LIGO, because otherwise the $k^3$ growth (relative to a
scale-invariant spectrum) will lead to excessive black hole production on
somewhat shorter scales, and also enough short-scale gravitational waves to
disrupt nucleosynthesis \cite{ab}.  However, the requirement that we see the 
end of the dilaton phase is not too
unreasonable, since the LIGO sensitivity is not far from requiring that
$A_{{\rm T}}$ be of order unity, a natural condition for string
effects to become important.

We would then require an abrupt turn over in the spectrum to avoid large
perturbations on shorter scales.  Such behaviour has in fact been found in a
toy model \cite{Franco}.  We therefore are in effect assuming a `minimal'
scenario, where it is assumed that significant perturbations are only 
generated in the
dilaton phase.\footnote{The expression for the perturbations also assumes
zero dilaton mass.  Since this is disallowed in the present universe, it is
usually argued that the dilaton acquires a mass at a relatively low energy
scale such as the supersymmetry scale.}  

If the expansion factor during the string phase exceeds $10^9$, the frequency
band accessible to the LIGO configuration would correspond to modes that went
beyond the Hubble radius during this phase rather than the dilaton phase.
The dynamics of this string phase where non-perturbative corrections are
expected to become important is extremely uncertain.
Gasperini \cite{gasalpha} has shown that if the space--time curvature and
kinetic energy of the dilaton field remain constant, the first--order
corrections in the inverse string tension do not significantly affect the
time evolution of the tensor perturbations, although the tilt of the spectrum
may deviate from three due to the unknown behaviour of the scale factor.

Maggiore and Sturani \cite{mags} have attempted to describe the evolution of
scalar perturbations through this era.  In practice our calculation of the
formation of black holes from density perturbations is not very sensitive to
the precise tilt of the spectrum and merely assumes that the perturbations
are growing towards a maximum at the scale $k_{{\rm s}}$.  Similarly the
relation between $A_{{\rm T}}^2$ and the logarithmic density $\Omega_{\rm
gw}$ on a given scale used in Eq.~(\ref{ratio}) does not depend on the
spectrum.  However we have assumed that the ratio $\epsilon$ is exactly $3$
(or less than $3$ if we allow non-adiabatic perturbations).  In all other
known inflationary scenarios the effective value of $\epsilon$ is less than 
$3$ and
it is tempting to conjecture that $3$ is the maximum possible value in any
inflationary scenario.  If so, the maximum density of gravitational waves
compatible with black hole limits would remain that given by
Eq.~(\ref{Omegagw}).

To produce $\epsilon>3$ requires that we suppress scalar
perturbations while still generating tensor perturbations. This seems
to be difficult in standard theories of inflation, where one will always
get perturbations in the field which controls the duration of
inflation, but in the absence of any specific calculation for the
perturbations in the string phase we cannot directly constrain the
gravitational wave spectrum in terms of the scalar perturbations.

Finally, it is worth remarking that larger black holes with masses in the
range $10^{-3}M_{\odot} \le M \le 1M_{\odot}$ could form by the mechanism
outlined above, if the string phase is of the correct duration to place the
end of the dilaton phase at the appropriate wavelength.  This would have
important implications for interpreting the microlensing events observed in 
our galaxy.

\subsection{Reheating}

The relation between modes leaving the horizon at the end of the dilaton
phase and their comoving scale in the radiation-dominated era depends not
only on the duration of the string phase, but also the reheat temperature at
the start of the hot big bang.  If reheating after the pre--big bang era is
due to the decay of weakly coupled massive particles produced at the end of
the string era, then the initial Hubble rate at the start of the hot big bang
could be well below the string scale $M_{\rm st}$.  The equation of state of
an extended phase dominated by a massive scalar field undergoing coherent
oscillations is effectively that of a pressureless fluid and this implies
that $aH \propto t^{-1/3}$ \cite{turnerosc}.  This may reduce the required
$10^9$ expansion of the string phase.  For example, if the universe is
dominated by such a field between energy scales $10^{18}{\rm GeV}$ and 
$10^9{\rm GeV}$, the
inflationary expansion during the string phase should be $10^6$ to place the 
end of the dilaton phase in the required range.

Black hole formation is not suppressed on frequencies above $f_{\rm s}$ if
the spectrum of scalar perturbations generated during the string phase is
flat or continues to grow on small scales.  An extended mass spectrum of
primordial black holes may form if the spectrum is precisely flat
\cite{carr}.  However, even a very small increase towards smaller scales
implies that the mass spectrum will be dominated by the smallest black holes.
This case
may have interesting consequences for string cosmology.  Black holes with
masses as small as $M = {\cal{O}} (m_{\rm Pl}^2/M_{\rm st})$ could then form,
where $m_{\rm Pl} \approx 10^{-5}{\rm g}$ is the Planck mass, and in most
supersymmetric grand unified theories, $10^{-2}< M_{{\rm st}}/m_{{\rm Pl}} <
10^{-1}$ \cite{stringscale}.  The only observational constraint below
$10^4\,{\rm g}$ arises if black holes leave behind stable Planck mass relics
in the final stages of their evaporation, but this now seems unlikely in view
of the recent developments in the understanding of black hole evaporation in
string theory (for a recent review, see, e.g.~Ref.~\cite{evap}).  The copious
production and rapid evaporation of black holes on these extremely small
scales then provides a natural mechanism for black hole reheating of the
universe after the string phase has ended \cite{cgl,bcl,garcia}.

\subsection{Thermal Inflation}

It is known that late entropy release, for instance from the decay of 
long-lived massive particles or evaporation of mini black holes, could 
suppress the present density of both black holes and gravitational waves 
\cite{BGV}.  However the constraint
on the scalar perturbation amplitude is relatively insensitive to the number
density of black holes, so late entropy release could only tighten the upper
limit on the maximum amplitude of gravitational waves.

An extreme version of late entropy release is a second, relatively 
short, period of inflation known as `thermal inflation' \cite{stewart}.  
Thermal inflation has been
proposed within the context of supersymmetric theories as a solution to a
generic problem of inflationary models, such as the pre--big bang scenario,
that have high reheat temperatures.  This problem arises because moduli
fields can come to dominate the universe before the onset of nucleosynthesis.
Thermal inflation resolves this problem by diluting the moduli fields' energy
density by a sufficient factor.  If $a_{{\rm i}}$ and $a_{{\rm f}}$ denote
the scale factors at the onset and end of thermal inflation respectively,
then, like other massive relics, the density of black holes is diluted by a 
factor $(a_{{\rm f}}/a_{{\rm i}})^3$ relative to the radiation produced at
the end of thermal inflation.  

Thermal inflation is driven by a scalar field with vacuum expectation value
$M \gg 10^3$GeV and mass corresponding to the supersymmetry scale, $m \approx
10^3$ GeV.  Typically, it begins when the temperature falls below $T \approx
(mM)^{1/2}$ and ends when $T \approx m$, so the expansion factor is $a_{{\rm
f}}/a_{{\rm i}} \approx (M/m)^{1/2}$. 
Successful nucleosynthesis requires $M \le 10^{14}$ GeV \cite{barreiro} and
this implies that thermal inflation must begin below about $10^8$ GeV. 

Equation (\ref{fs}) requires modification if thermal inflation occurs, and 
one finds that the modes relevant to LIGO are within the Hubble radius 
throughout thermal 
inflation. This dramatically alters the density of gravitational waves on 
these
wavelengths. Once gravitational waves have re-entered the Hubble
length their energy density evolves like ordinary radiation. Although
it is not diluted relative to other radiation in the standard
radiation dominated era, it is diluted relative to the total energy
density during thermal inflation. Thus, for scales which remain within
the horizon throughout a period of thermal inflation, the energy
density is diluted by a factor $(a_{{\rm i}}/a_{{\rm 
f}})^4\approx(m/M)^2\sim10^{-22}$. Because we require that the
initial amplitude $A_{{\rm T}}^2$ is less than unity, this dilutes the
intensity of gravitational waves to way below the LIGO sensitivity.

This is quite a powerful and model-independent conclusion. 
Thermal inflation makes it impossible for LIGO to see a stochastic 
gravitational wave background generated in the very early universe.

\section{Summary}

In the simplest pre--big bang scenario, a detection of gravitational waves at 
a high level implies that there should also be significant black hole 
production. There are many ways in which the 
simplest scenario may need amendment, such as nonadiabatic perturbations or 
late entropy production, and these changes reduce the maximum allowed 
gravitational wave 
amplitude. Therefore, if one were to 
detect gravitational waves at a high level, above that given by 
Eq.~(\ref{Omegagw}), without detecting black holes, it would suggest they 
were not produced in a dilaton driven pre--big bang phase. 

Alternatively, one might detect both, but with the gravitational waves well 
below the result of Eq.~(\ref{Omegagw}). That would give an estimate of the 
importance of the various additional effects which would need to be 
incorporated into the scenario. However, if gravitational waves are not 
detected, this could be for any of several reasons and would not say anything 
much about the pre--big bang scenario.

To conclude, in the pre--big bang cosmology there are prospects for detection 
of both gravitational waves and black holes. Detection of the two in concert 
would provide strong supporting evidence for the pre--big bang scenario. 

\section*{Acknowledgments}

E.J.C.~and J.E.L.~were supported by PPARC and A.R.L.~by the Royal
Society. We thank M.~Giovannini, J.~Maharana and G.~Veneziano 
for useful discussions.  

 
\end{document}